\documentclass[conference]{IEEEtran}
\IEEEoverridecommandlockouts

\usepackage{amsmath,amssymb,amsfonts}
\usepackage{algorithmic}
\usepackage{graphicx}
\usepackage{textcomp}
\usepackage{xcolor}
\usepackage{booktabs}

\usepackage[style=ieee,giveninits=true]{biblatex}
\begin{document}

\title{Using Ear-EEG to Decode Auditory Attention in Multiple-speaker Environment\\

\thanks{This work is supported by the National Key Research and Development Program of China (No.2021ZD0201503), a National Natural Science Foundation of China (No.12074012) and the High-performance Computing Platform of Peking.}
}
\DeclareRobustCommand*{\IEEEauthorrefmark}[1]{%
  \raisebox{0pt}[0pt][0pt]{\textsuperscript{\footnotesize #1}}%
}
\author{
\IEEEauthorblockN{Haolin Zhu\IEEEauthorrefmark{1}\IEEEauthorrefmark{3}, 
    Yujie Yan\IEEEauthorrefmark{2}\IEEEauthorrefmark{3}, 
    Xiran Xu\IEEEauthorrefmark{1}\IEEEauthorrefmark{3}, 
    Zhongshu Ge\IEEEauthorrefmark{1}, 
    Pei Tian\IEEEauthorrefmark{1}, 
    Xihong Wu\IEEEauthorrefmark{1}\IEEEauthorrefmark{3},
    Jing Chen\IEEEauthorrefmark{1}\IEEEauthorrefmark{2}\IEEEauthorrefmark{3}} \\
    \IEEEauthorblockA{
    \IEEEauthorrefmark{1} Speech and Hearing Research Center, School of Intelligence Science and Technology, Peking University\\
    \IEEEauthorrefmark{2} Center for BioMed-X Research, Academy for Advanced Interdisciplinary Studies, Peking University\\
    \IEEEauthorrefmark{3} National Key Laboratory of General Artificial Intelligence, Peking University\\
    janechenjing@pku.edu.cn
    }
}
\maketitle

\begin{abstract}
Auditory Attention Decoding (AAD) can help to determine the identity of the attended speaker during an auditory selective attention task, by analyzing and processing measurements of electroencephalography (EEG) data. Most studies on AAD are based on scalp-EEG signals in two-speaker scenarios, which are far from real application. Ear-EEG has recently gained significant attention due to its motion tolerance and invisibility during data acquisition, making it easy to incorporate with other devices for applications. In this work, participants selectively attended to one of the four spatially separated speakers' speech in an anechoic room. The EEG data were concurrently collected from a scalp-EEG system and an ear-EEG system (cEEGrids). Temporal response functions (TRFs) and stimulus reconstruction (SR) were utilized using ear-EEG data. Results showed that the attended speech TRFs were stronger than each unattended speech and decoding accuracy was 41.3\% in the 60s (chance level of 25\%). To further investigate the impact of electrode placement and quantity, SR was utilized in both scalp-EEG and ear-EEG, revealing that while the number of electrodes had a minor effect, their positioning had a significant influence on the decoding accuracy. One kind of auditory spatial attention detection (ASAD) method, STAnet, was testified with this ear-EEG database, resulting in 93.1\% in 1-second decoding window. The implementation code and database for our work are available on GitHub: https://github.com/zhl486/Ear\_EEG\_code.git and Zenodo: https://zenodo.org/records/10803261.
\end{abstract}


\section{Introduction}
In a cocktail party scenario where multiple speakers are talking simultaneously \cite{cherry1953some}, listeners are able to attend selectively to the target speech while ignoring others. Neuroscientific studies suggest that the attentional focus can be identified through listeners’ EEG, as selective attention has been found to enhance cortical tracking of the attended speech and suppress synchronization of the ignored speech \cite{ding2012emergence,mesgarani2012selective,o2015attentional}. The corresponding technology developed to decode the attended object with EEG is named auditory attention decoding (AAD), which is potentially applicable to the realization of cognitively controlled, or neuro-steered, hearing equipment.

Most studies on AAD are based on scalp-EEG \cite{xu2024densenet,xu2024convconcatnet,qiuexploring,yan_auditory_2024}. For unobtrusive EEG acquisition, small and near-invisible approaches are preferred not to disturb natural social interaction \cite{looney2012ear,debener2015unobtrusive,mirkovic2016target,holtze2022ear,somon2022benchmarking,bleichner2016identifying}. This demand has led to the development of in-ear EEG \cite{looney2012ear}, and around-the-ear EEG solutions \cite{debener2015unobtrusive,mirkovic2016target,holtze2022ear,somon2022benchmarking,bleichner2016identifying}, where electrodes are placed inside the outer ear canal or around the ear, respectively. Considering the larger distance between electrodes compared to in-ear EEG, around-the-ear EEG was used for this study \cite{bleichner2017concealed}. The cEEGrid is one around-the-ear EEG solution which is a C-shaped flex-printed sensor array comprising 10 electrodes. Previous studies \cite{mirkovic2016target,holtze2022ear,bleichner2016identifying} have confirmed the feasibility of utilizing cEEGrids to decode the attended speaker. For example, Mirkovic et al. further elucidated that in more complex processes, such as two-speaker scenarios, these signals manifested discernible traces in the cEEGrids EEG recordings, resulting in a decoding accuracy of 69.3\% \cite{bleichner2016identifying}. Holtze et al. performed artifact correction on the ear-EEG, resulting in the 72.13\% average decoding accuracy reached with SR in the two-speaker scenario in the 60-second decision window \cite{holtze2022ear}. Thornton et al. W. Nogueira et al. demonstrated the suitability of cEEGrids for cochlear implant users \cite{mirkovic2016target}. Due to their unobtrusive and concealed design, cEEGrids offer more promising potential applications compared to scalp-EEG. 

In scenarios where speakers are spatially separated, auditory spatial attention detection \cite{geirnaert2020fast,geirnaert2020eeg,geirnaert2021riemannian,li2022esaa,su2022stanet,deckers2018eeg,zhang2023learnable} have been developed to decode the attended spatial location,  providing an alternative approach to decoding the attended object, in addition to SR.  Due to its high accuracy on short time-scales (within 1–5 seconds), ASAD is well-suited for real-time applications. Among the existing ASAD methods, DNN-based approaches \cite{su2022stanet,zhang2023learnable} have demonstrated superior performance over traditional decoding strategies due to their exceptional capabilities in feature extraction and nonlinear representation. The integration of attention mechanisms into ASAD \cite{geirnaert2020fast,zhang2023learnable} has marked a recent innovation. This has enabled STAnet to attain a decoding accuracy of 90.1\% \cite{li2022esaa} within 1-second decision windows on scalp-EEG KUL datasets \cite{biesmans2016auditory}. However, currently, it’s still unclear whether those ASAD methods for EEG can be extended to ear-EEG data.

In this work, to determine if the identity of the attended speaker can be captured with ear-EEG in a more realistic scenario, participants were instructed to direct their attention to one of the four spatially separated speakers, with one speaker serving as the target speaker and the other three the interference and their ear-EEG were recorded. The forward temporal response functions (TRFs) analysis and the backward SR method are employed to analyze the dynamic properties of cortical envelope tracking activities in this complex environment and explore the feasibility of SR-based AAD on ear-EEG data when more than one interference speeches are presented. As an alternative to SR, ASAD using STAnet may be used to further get higher performance. The contributions of this work are as follows: 1) We verified the feasibility of using ear-EEG for auditory attention decoding under the condition of four spatially separated speakers. 2) We compared the decoding accuracy of ear-EEG to scalp-EEG collected simultaneously and investigated the influence of electrode placement and quantity. 3) We released the first ear-EEG database in public as a benchmark for common research and discussion. 
\section{METHODS}

\subsection{Subjects}

Sixteen subjects (6 females, age range: 19–27 years) from Peking University with normal hearing participated in the experiment. All subjects were Mandarin-native speakers who reported no medical history of brain injury or cognitive deficits. Before the experiments, subjects were briefed on the procedure and the objectives, given informed consent in accordance with a protocol approved by the Peking University Institutional Review Board.

\subsection{Stimuli and experimental procedure}
The audio stimuli used in this experiment were from the Chinese speech materials (\textit{Twenty Thousand Leagues under the Sea by Jules Verne}) detailed in previous work \cite{fu2020congruent,fu2019congruent,yan_auditory_2024}. The preprocessing of the stimuli was the same as in previous work \cite{yan_auditory_2024}.

In the experiment, four speech segments in a combination were presented respectively through four loudspeakers (Dynaudio BM 6A)  positioned at +30°, -30°, +90° and -90°, aligned with the height of the subjects' ears, at a sound pressure level of 55 $dBLA_{eq}$ \cite{yan_auditory_2024,fu2019congruent}. The four loudspeakers were positioned within the 1.6m half-circle area with the subject’s head in the center of the half-circle. The subjects were instructed to follow the speech from a target direction while ignoring the others. During the stimulus presentation, subjects were instructed to maintain visual fixation on the white crosshair displayed on the computer screen positioned right in front of them and keep their head fixed. After each trial, subjects were required to choose the accurate answers to 4 questions regarding the content of the attended speech.

\subsection{EEG data acquisition and preprocessing}
During the experiment, subjects were seated in a comfortable chair in an anechoic room. EEG was recorded with two kinds of recording devices. Previous research has demonstrated that using two cEEGrids does not disrupt the signal quality obtained from scalp-EEG recordings \cite{yan_auditory_2024}. 
\begin{figure}[t]
\centerline{\includegraphics[width=\linewidth]{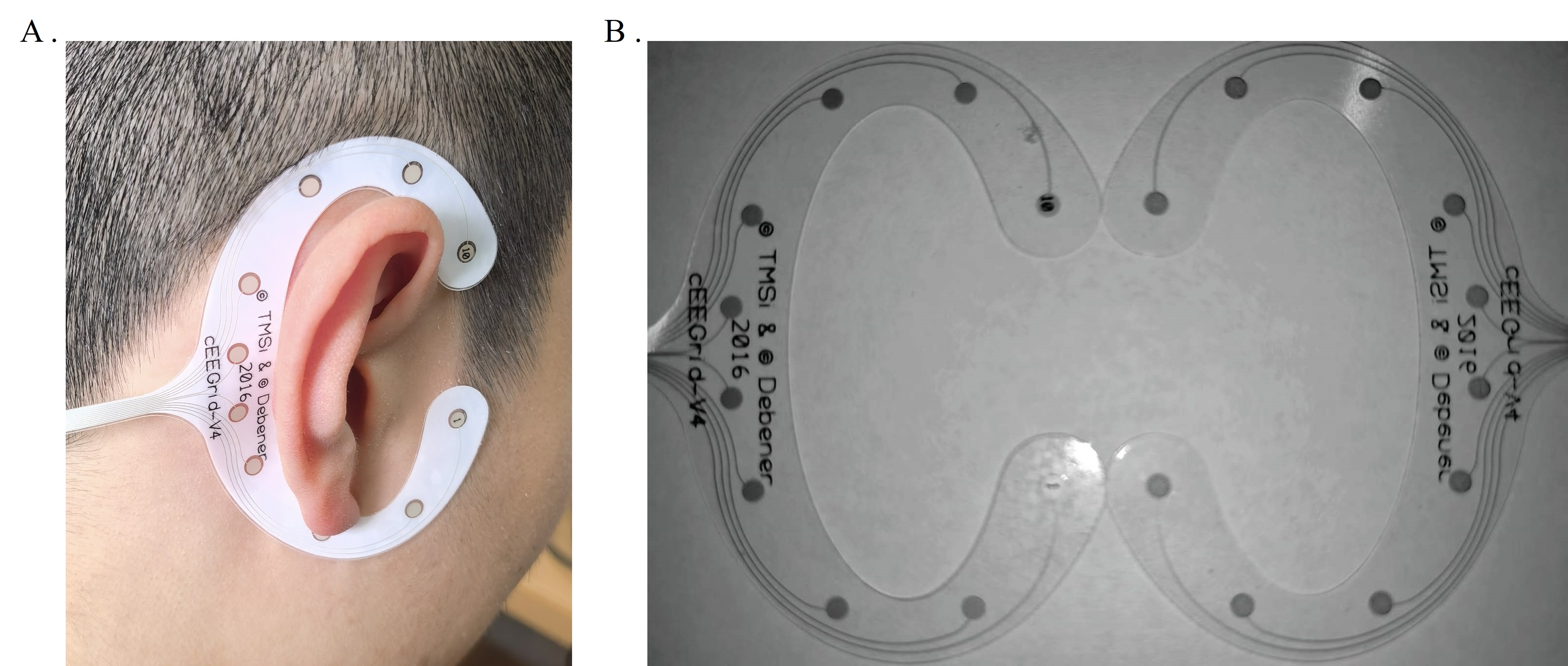}}
\caption{cEEGrid: A C-shaped electrode array is affixed around the ear using a double-sided adhesive. Each cEEGrid consists of 10 electrodes. A. Picture of a cEEGrid placed around the right ear. B. The electrode positions of two cEEGrids for left and right ear.
 }
\label{fig1}
\end{figure}
Continuous ear-EEG data were recorded with a SAGA 32+/64+ amplifier (TMSi, Oldenzaal, the Netherlands) and two cEEGrids \cite{debener2015unobtrusive}, acquired by Polybench 1.34 software (TMSi, Oldenzaal, the Netherlands). After skin preparation with an abrasive gel and alcohol, a small amount of electrolyte gel (GT5, GREENTEK, China) was applied to the electrodes and the two cEEGrids (shown in Fig.\ref{fig1}.) were placed around both sides of the ear. An additional electrode attached to the wrist was used as the ground. Ear-EEG recordings were re-referenced to a common average reference, sampled at 500 Hz and stored for offline analysis.

The scalp-EEG database was the four-talker EEG database in \cite{yan_auditory_2024}. This database contains the EEG recordings from 16 normal hearing subjects and 40 minutes for each of them. The EEG signal was recorded by a 64-channel EEG NeuSen Recorder (Neuracle, China) at a sampling rate of 1000 Hz. More details including the data preprocessing are available in \cite{yan_auditory_2024}.

For SR, both EEG data were band-pass filtered between 2 and 8 Hz, since the low-frequency ($<$8 Hz) neural activity in the auditory cortex was reported phase locked to speech envelopes \cite{cherry1953some,ding2012emergence,mesgarani2012selective} and subsequently baseline corrected and down-sampled to 64 Hz. The aforementioned process was performed using the EEGLAB toolbox on MATLAB \cite{delorme2004eeglab}.

\subsection{TRFs estimation}
The TRFs estimation methods employed in this study were consistent with those described in previous research \cite{fu2020congruent,yan_auditory_2024}. TRFs analysis was conducted using the mTRF toolbox \cite{crosse2016multivariate}. For speech temporal envelope $s(t)$ sampled at discrete time $t (t = 1, \cdots, T)$ and the corresponding EEG $r(t, n)$ recorded at channel $n (n = 1, \cdots, N)$, suppose that a set of spatio-temporal filters (i.e., the channel-specific TRFs) $w(\tau, n)$ could map from $s(t)$ to $r(t,n)$ in a convolutional way, as in equation \ref{equation:eq1}:
\begin{align}
  \hat{r}(t, n)=\sum_{\tau} \mathrm{w}(\tau, \mathrm{n}) \mathrm{s}(t-\tau)
  \label{equation:eq1}
\end{align}
where $\hat{r}(t, n)$ represents estimated EEG. The TRFs coefficients were calculated using the reverse correlation method with the ridge regression to solve the ill-posed problems and overfitting. A range of time delays, spanning from -50ms before the stimulus to 450ms after the stimulus, was employed. 

\subsection{Speech stimulus reconstruction}
The speech temporal envelope extraction and SR procedure were the same as previous work \cite{o2015attentional}. Briefly, a set of filters was used to map EEG to speech envelopes. Let's $s(t)$ be the temporal envelope of a certain speech stimulus sampled at discrete time $t (t = 1, \cdots, T)$, and $r(t, n)$ is the corresponding recorded neural at EEG channel $n (n = 1, \cdots, N)$. The spatio-temporal filter $g(\tau, n)$, also termed ``decoder'', represent the linear backward mapping from $r(t, n)$ to $s(t)$, as in equation \ref{equation:eq2}:
\begin{align}
\hat{s}(t)=\sum_{n} \sum_{\tau} r(t+\tau, n) g(\tau, n)
  \label{equation:eq2}
\end{align}
To investigate the impact of electrode layout, three different layouts (shown in Fig. 4A.) were utilized with scalp-EEG data, while maintaining an identical cluster size to the number of ear-EEG channels (20). 

\subsection{ASAD}
\begin{figure}[t]
\centering
\includegraphics[width=\linewidth]{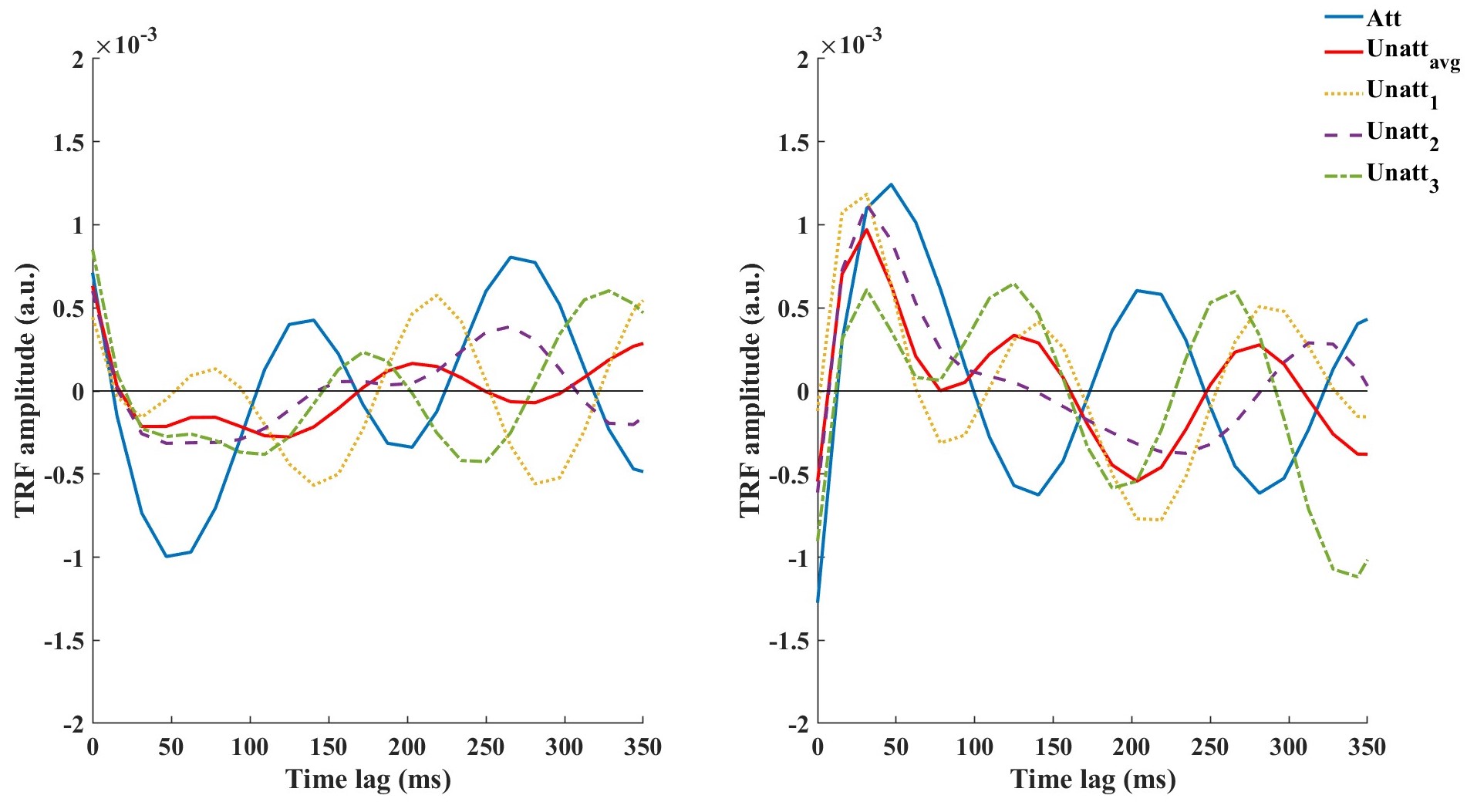}
\caption{Ear-EEG data from two electrodes positioned on two cEEGrids were selected for TRF analysis. The attended TRFs averaged over the other three unattended TRFs. TRFs of the other three unattended speeches are plotted as colored dashed lines.}
\label{fig:trf}
\end{figure}

Given the difficulties encountered in acquiring clean speech data in real-world scenarios, coupled with the requirements of maintaining promptness and precision, the implementation of SR is frequently beset with challenges. An end-to-end network, STAnet, has previously demonstrated its robustness on low-density EEG data by selecting 16 channels from the available 64 channels and achieved an average decoding accuracy of 75.4\% (DTU: 66.4\%, KUL: 84.4\%) in the two-speaker environment. 

Experiments were conducted on four ASAD models to validate the feasibility of the ASAD method on the current database. The first model (CNN-baseline) was a CNN model as the same as \cite{deckers2018eeg}. The second model, SAnet, removes the temporal feature representation module from STAnet, and the third model, TAnet, removes the spatial attention mechanism from the spatial feature representation module of STAnet. The last model remains the original STAnet. More details in models are referred to \cite{su2022stanet}. 

\section{RESULTS AND DISCUSSION}
\subsection{TRFs estimation}
 To assess the validity of ear-EEG in a four-talker environment with spatial separation, average temporal response functions (TRFs) for both attended and unattended speech were estimated across all subjects and trials. As depicted in Fig. \ref{fig:trf}, ear-EEG data from two electrodes positioned on two cEEGrids were selected for TRFs analysis. The results display the averaged TRFs calculated separately for each of the three unattended speeches, with the TRFs of the remaining three unattended speeches also represented in the figure as colored dashed lines. It was observed that the TRFs response to attended speech was higher than that to unattended TRFs, consistent with previous findings in four-talker scenarios \cite{yan_auditory_2024}. 

\begin{figure}[!t]
\centering
\includegraphics[width=\linewidth]{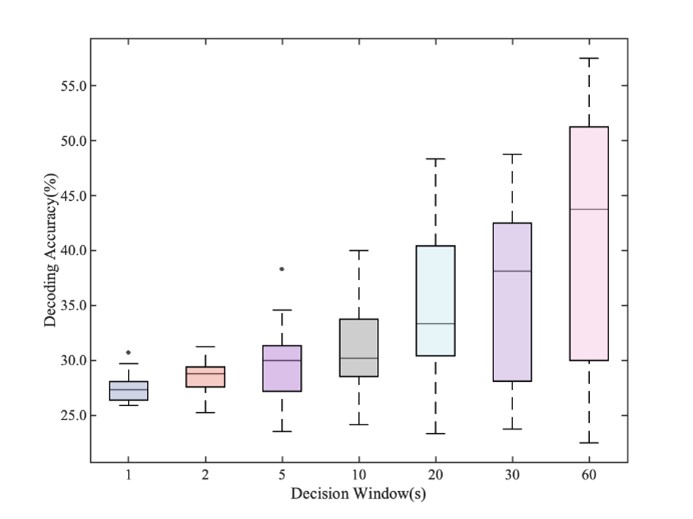}
\caption{The result of SR in different decoding window lengths.}
\label{fig:sr}
\end{figure}

\subsection{Stimulus reconstruction }
\begin{figure*}[t]
  \centering
  \includegraphics[width=1.0\textwidth]{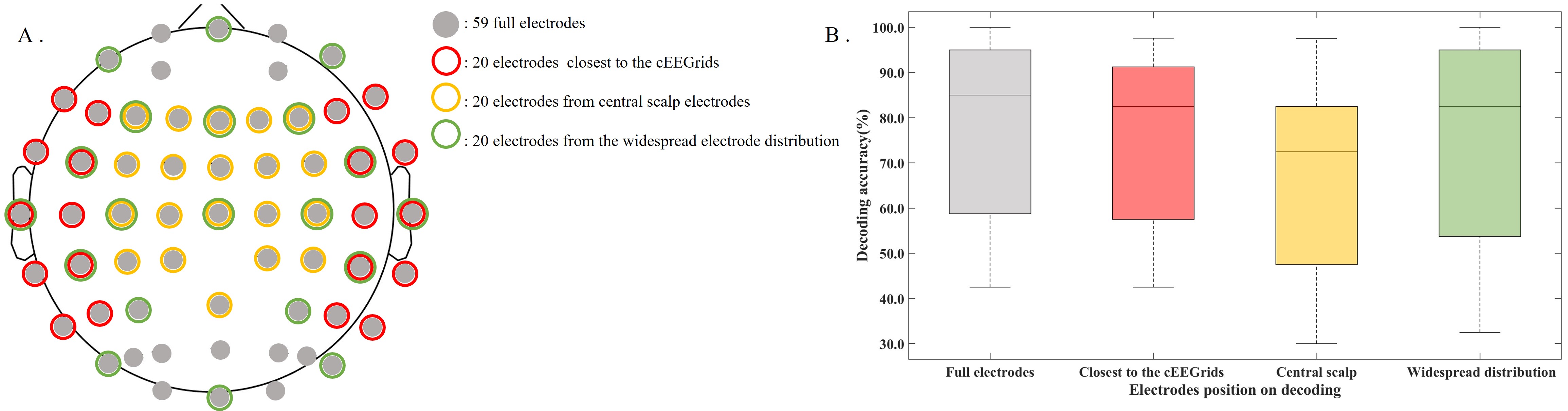}
  \caption{A. Four different electrode layouts: Gray dot: 59 full electrodes. Red circle:  20 electrodes closest to the cEEGrids. Orange circle: 20 electrodes from central scalp electrodes. Green circle: 20 electrodes from the widespread electrode layout; B. Decoding accuracy for four different electrode layouts}
  \label{Figure 4}
\end{figure*}
In a more realistic environment, noise-robust of attended speech was discovered, where the target speech remained highly intelligible despite signal distortions. The decoding accuracy averaged across subjects for each decision window is shown in Fig. \ref{fig:sr}. The accuracies were 27.5\% (SD:1.3\%), 28.5\% (SD: 1.6\%), 29.8\% (SD: 3.7\%), 31.1\% (SD: 4.5\%), 35.0\% (SD: 7.1\%), 36.4\% (SD: 1.9\%), 41.3\% (SD: 11.0\%) for the decision window length of 1, 2, 5, 10, 20, 30, 60 seconds, respectively. The accuracies were significantly higher than the chance level, suggesting the feasibility of SR-based AAD with ear-EEG in multiple $(>1)$ interference auditory scenarios.

The results of SR using scalp-EEG were the same as our previously reported in \cite{yan_auditory_2024}, and this work investigates the influence of electrode placement and quantity as discussed in \cite{mirkovic2016target}, shown in Fig. \ref{Figure 4}. B. The 59 available scalp channels (1 reference (nose-tip) and 4 bipolar electrooculogram electrodes were not utilized) produced a decoding accuracy of 77.50\% \cite{yan_auditory_2024}. By reanalyzing our previously published scalp-EEG data using around-the-ear 20 channels only, we achieved a decoding accuracy of 75.47\% (SD: 21.06\%). Compared to layouts closest to the cEEGrids, a cluster with 20 channels covering only central scalp channels yielded the lowest performance at 67.50\%. Reducing the number of channels to 20 while maintaining an even distribution across the scalp resulted in a slight decrease in accuracy, achieving 75\% (SD:23.84\%) in the 60-second decision window. 

Compared to scalp-EEG with 59 channels, the accuracy of SR using ear-EEG was compromised by the decreased electrode count and concentrated distribution, potentially leading to a loss of spatial information. The reduction of channels made the accuracy decrease less obvious seems that the main factor contributing to the lower performance of ear-EEG compared to scalp-EEG was the spatial location of electrode placement. Compared with the coverage from the central channels, a better result was achieved by using data collected from the 20 scalp channels closest to the cEEGrids, which illustrated the feasibility of collecting signals from specific regions for efficient auditory attention decoding with fewer electrodes.
\subsection{ASAD}
For ASAD, the four aforementioned models achieved an average decoding accuracy of 84.5\% (SD: 6.5\%), 92.4\% (SD: 3.1\%), 92.9\% (SD: 2.5\%), and 93.1\% (SD: 3.0\%) in 1-second decision window. The accuracies of the four models across all subjects were notably superior to those obtained using SR-based AAD. SAnet, TAnet, and STAnet each achieved a significantly higher average decoding accuracy compared to the CNN-baseline model, suggesting the effectiveness of introducing the attention mechanism to the ASAD method. However, there was no significant difference in accuracy between these three models (SANet: 92.4\%, TANet: 92.9\%, STAnet: 93.1\%) showed no significant difference. This observation may be attributed to the robust capacity of SAnet and TAnet to extract spatio-temporal features, potentially leading to a saturation effect in accuracy improvement. Nonetheless, recent studies \cite{rotaru2024we,qiu2024enhancing,xu2024beware,qiu2024streamaad} have indicated that EEG data from the same trial may contain similar, trial-specific features, which could cause the network to infer trial-related information rather than accurately decode attention, thereby inflating the results.

 Consistent with previous findings in \cite{nogueira2019decoding,bleichner2016identifying,holtze2022ear,mirkovic2016target}, the results demonstrate the strong performance of a low-density EEG system (i.e., cEEGrids) for decoding. This highlights the potential of using ear-EEG to decode the attended object in real-life scenarios.

\subsection{Limitations}
The experimental paradigm presented here was limited to a four-speaker scenario. In real-world settings, the complexity increases as environmental sound sources multiply, transforming the task into more than a basic four-way classification problem, which can significantly impact decoding performance. Additionally, the experiments were conducted in an anechoic room. In real environments, reverberation could reduce the cortical tracking precision of speech, weakening the distinction between neural responses to attended and unattended speakers and adding further challenges to the decoding process. Furthermore, the interpretability of the ASAD algorithm remains relatively low. Future work will focus on developing more practical and accurate applications of ear-EEG that better reflect real-world conditions.
\section{CONCLUSIONS}
The present work suggests that it is feasible to use ear-EEG to auditory attention decode (AAD) in a four-speaker scenario. Through stimulus reconstruction (SR) and temporal response functions (TRFs), the results show that neural activity recorded with ear-EEG demonstrates more tracking to the attended speech envelope compared to the unattended ones, making the decoding accuracy significantly higher above the chance level. This work represents the first application of the auditory spatial attention detection (ASAD) algorithm on ear-EEG for AAD tasks. The decoding results of STAnet indicate the potential of ear-EEG in decoding attended spatial locations within a short decision window. By using concurrent ear-EEG and scalp-EEG recordings, this work provides evidence that the decoding accuracy is less influenced by the quantity of electrodes and more influenced by the spatial distribution of electrodes.
\renewcommand{\bibfont}{\footnotesize}
\printbibliography

\end{document}